\documentclass[12pt]{article}
\begin{document}

\title{
 Thermodynamics of Dark Energy}
\author{
Neven Bili\'c
 \\
 Rudjer Bo\v{s}kovi\'{c} Institute, 10002 Zagreb, Croatia \\
E-mail: bilic@thphys.irb.hr \\
}
\maketitle



\begin{abstract}
  Thermodynamic properties of 
dark energy are discussed assuming that dark energy is described in terms
of a selfinteracting complex scalar.
We first show that, under certain assumptions,
 selfinteracting complex scalar field theories
are equivalent to purely kinetic $k$-essence models.
Then we analyze the themal properties of $k$-essence
and in particular we show
that dark-energy in the phantom regime
does not necessarily yield negative entropy.
\end{abstract}
\maketitle                   






\section{Introduction}
A motivation for and importance of studying the dark energy (DE) thermodynamics can be
summarized in the following questions: Does a DE fluid possess
thermal, besides hydrodynamical, 
properties such as temperature? What is the thermodynamical fate of the Universe?
Can DE cluster?
In a number of recent papers
\cite{bre,lim,gonz,izq,moh,set,wan,gong,san},
thermal properties of dark energy
have been discussed based on the assumption that
the dark energy substance is a thermalized
ensemble at certain temperature
with an associate  thermodynamical entropy.
Here we hope to shed some light on a rather
peculiar thermodynamic related properties of DE 
often discussed  in the literature. These may be illustrated
by a few statements taken from recent papers:
DE becomes hotter if it
undergoes an adiabatic expansion \cite{lim}.
Phantom DE violates the null energy condition  and hence
either its entropy  or its temperature must be negative \cite{moh,gong}.
A negative temperature implies either that the phantom should be quantized or
that its space-time should be Euclidean \cite{gonz}.

The essential feature of DE is that its pressure must be negative in order
to reproduce the accelerated expansion of the Universe.
Hence, the DE equation of state violates the strong energy condition.
The simplest DE model, the {\em cosmological constant}, is basically the 
vacuum energy, with the equation of state $p=-\rho$. As a consequence, in the
course of the Universe evolution, the vacuum energy density remains
constant and its thermal properties are trivial.
In other DE models with more complicated equation of state (EOS)
the energy density varies with time.
A  number of models, such as {\em quintessence} \cite{peeb8} and 
{\em k-essence} \cite{arm2}, 
are based on scalar field theories. To this class  belong also the so called
{\em quartessence} models \cite{mahl7}, where the term quartessence was invented to denote
the unification of dark energy and dark matter. A very popular class
of models is the {\em phantom DE}, 
the EOS of which violates even the null energy condition (NEC), 
i.e, in these models $(p+\rho) <0$.
For an excellent recent review of DE models, see \cite{cop}).

Dark energy  is usually described by an EOS in the barotropic form,
i.e., in the form $p=p(\rho)$. Equivalently, one may define a field theory
Lagrangian, in which dark energy is described in terms of
 a classical selfinteracting field
coupled to gravity.
 Then, the EOS may  be deduced from the
energy-momentum tensor  obtained from  variational principle.
From the EOS alone
it is not possible to uniquely determine
the thermodynamic properties of a system.
One simple example is the EOS
$p=\rho/3$ which may describe a massless boson gas at $T\neq 0$ (hence $S\neq 0$)
but also a massless degenerate Fermi gas at $T=0$ (hence $S=0$).
A similar situation occurs for any barotropic EOS.

The purpose of this paper is twofold.
First, we would like to demonstrate 
that, under reasonable assumption in the cosmological context,
the ghost condensate and purely kinetic $k$-essence models are
equivalent to standard, selfinteracting complex scalar field theories.
Second, we analyze the thermal properties of  a grand canonical DE gas
described by  purely kinetic k-essence with a chemical potential
that corresponds to a conserved charge.
The conserved charge $Q$ is related to the shift symmetry
$\phi \rightarrow \phi + {\rm const}$ which corresponds to to the
$U(1)$ symmetry of the equivalent complex scalar field theory.
Hence, the conserved quantity $Q$ corresponds to the usual $U(1)$ charge
of a complex scalar field.
In this way a consistent grandcanonical description of DE
involves the thermodynamic equations with two variables:
the temperature $T$ and chemical potential $\mu$.

\section{Thomas-Fermi correspondence}
\label{correspondence}

Consider the
Lagrangian
\begin{equation}
{\cal{L}}
 = \eta g^{\mu \nu} {\Phi^*}_{, \mu}
\Phi_{, \nu}
 -V(|\Phi|^{2}/m^2)
\label{eq1101}
\end{equation}
for  a complex scalar field
\begin{equation}
\Phi = \frac{\phi}{\sqrt{2}}\exp (-im \theta).
\label{eq1102}
\end{equation}
where the potential $V$ may contain the mass  term
$m^2 |\Phi|^2$ and higher order selfinteraction terms.
Here  $\eta=1$ for a canonical scalar field and $\eta=-1$ for a phantom.
The field $\Phi$ satisfies the Klein-Gordon equation of motion
\begin{equation}
\frac{1}{\sqrt{-g}} ( \sqrt{-g}  g^{\mu \nu} {\Phi}_{, \nu})_{,\mu}
 + \frac{dV}{d|\Phi|^2} \Phi=0
\label{eq2103}
\end{equation}
If the space-time variations of $|\Phi|$ are
small on the scale smaller than $m^{-1}$, i.e.
assuming
$\phi_{, \mu} \ll m \phi$ ,
then the
Thomas-Fermi (TF) approximation \cite{par18,bil4}
applies and the derivatives of $\Phi$ may be calculated as
$\Phi_{, \mu} = -i m\Phi \theta_{,\mu}$.
 Hence, neglecting
$g^{\mu \nu} {\phi}_{, \mu}
\phi_{, \nu}/m^2$, the Lagrangian
(\ref{eq1101}) may be written as
\begin{equation}
{\cal{L}}_{\rm TF}/m^4=
   XY
 -U(Y)\, ,
 \label{eq1103}
\end{equation}
where we have introduced the abbreviations
\begin{equation}
X = g^{\mu \nu} {\theta}_{, \mu}
\theta_{, \nu}\, ;
 \;\;\;\;\;\;
Y=\eta\frac{\phi^2}{2m^2} \, ,
\label{eq1104}
\end{equation}
and the dimensionless potential
\begin{equation}
U(Y)=\frac{1}{m^4} V(\eta Y).
 \label{eq1303}
\end{equation}
The equations of motion for the field $\phi$ and $\theta$ are
\begin{equation}
X- U_Y=0 \,,
\label{eq1204}
\end{equation}
\begin{equation}
(Y g^{\mu \nu} {\theta}_{, \nu})_{;\mu}=0,
\label{eq1206}
\end{equation}
where $U_Y=dU/dY$.
The set of equations
(\ref{eq1204}), (\ref{eq1206}) is basically 
a reduced Klein-Gordon equation (\ref{eq2103}).
Equation (\ref{eq1104}) implies
$\eta Y>0$.
Assuming $X>0$ and hence 
\begin{equation}
U_Y>0,
\label{eq1306}
\end{equation}
the field $\theta$ may be treated as a
velocity potential for the fluid 4-velocity
\begin{equation}
u^{\mu} = g^{\mu \nu} \theta_{,\nu} / \sqrt{X} \, ,
\label{eq1215}
\end{equation}
satisfying the normalization condition
$u_{\mu} u^{\mu}$ = 1.
As a consequence,
 the energy-momentum tensor $T^{\mu \nu}$ obtained
from the Lagrangian (\ref{eq1103})
\begin{equation}
T_{\mu\nu}= 2\frac{\partial {\cal L}_{\rm TF}}{\partial X}\:
\theta_{,\mu}\theta_{,\nu}
-{\cal L}_{\rm TF}g_{\mu\nu}
\label{eq509}
\end{equation}
takes the perfect fluid
form, 
\begin{equation}
T_{\mu\nu}= (\rho+p) u_\mu u_\nu - p g_{\mu\nu}
\label{eq510}
\end{equation}
with the parametric equation
of state
\begin{equation}
\rho /m^4= Y U_Y+ U ,
 \hspace{1cm}   p/m^4 = Y
U_Y - U  .
\label{eq1216}
\end{equation}
A perfect fluid description applies
only if  (\ref{eq1306}) holds
which, generally, may not be true for the entire range $0\leq \eta Y \leq \infty$.
For a canonical field ($\eta=1$) the fluid is perfect
for those $Y$ for which $dV/d|\Phi|^2>0$. In contrast, a phantom field ($\eta=-1$)
behaves as a perfect fluid
when $dV/d|\Phi|^2<0$.

Owing to (\ref{eq1204}) and the obvious relation
$\partial{\cal{L}}_{\rm TF}/\partial X=m^4 Y$
 equation (\ref{eq1103}) may be written as a Legendre
transformation
\begin{equation}
W(X)+U(Y)=
   XY
  \label{eq1105}
\end{equation}
with 
\begin{equation}
X= U_Y
 \label{eq1406}
\end{equation}
\begin{equation}
Y= W_X
 \label{eq1106}
\end{equation}
where $W_X=dW/dX$.
Equation (\ref{eq1105}) with (\ref{eq1406}) and (\ref{eq1106}) defines an equivalence  relation.
We say that
potentials $W(X)$ and $U(Y)$ are {\em TF equivalent} to each other.
Given $U\equiv V(\eta Y)/m^4$,  the potential 
$W(X)$ can be found by solving (\ref{eq1406}) for $Y$
 and plugging the solution in (\ref{eq1105}).
More explicitly
\begin{equation}
W(X)= XU^{-1}_{Y}(X) -U(U^{-1}_{Y}(X)),
  \label{eq1405}
\end{equation}
where $U_Y^{-1}$ is the inverse function of $U_Y$.
Similarly, if $W(X)$ is known, the potential $U$ may be derived in the same way.

The potential $W(X)$ that is TF equivalent to $U(Y)$ describes a field theory with
the Lagrangian
\begin{equation}
{\cal L} = m^4 W(X)\, ;\; \;\;\;\; X \equiv g^{\mu \nu} \theta_{, \mu} \theta_{, \nu}
\label{eq1203}
\end{equation}
 which depends only on the derivatives of a scalar field 
$\theta$.
The equation of motion for $\theta$
\begin{equation}
(W_X g^{\mu \nu} \theta_{, \nu})_{;\mu}=0
\label{eq1304}
\end{equation}
is equivalent to (\ref{eq1206}).
However,
the field theories
described by the Lagrangians (\ref{eq1103}) and (\ref{eq1203}), respectively,
are equivalent
only at the classical level, since
the Lagrangian (\ref{eq1203}) is obtained 
from (\ref{eq1103}) by eliminating one degree of freedom with 
help of the equation of motion (\ref{eq1204}).
Quantum mechanically, these two field theories describe
different physics.

The energy-momentum tensor constructed
from the Lagrangian (\ref{eq1203})
is of the 
form (\ref{eq510}), with the parametric equation
of state
\begin{equation}
\rho/m^4 = 2X W_X -W,
 \hspace{1cm} p/m^4 = W.
\label{eq1316}
\end{equation}
This equation describes an ordinary fluid for 
$W_X>0$ and phantom for $W_X<0$.
The above mentioned equivalence between the two theories may be seen by noting that
equations (\ref{eq1216}) and (\ref{eq1316})  are
different parameterization of the same equation of state. This may be easily
verified by applying (\ref{eq1406}) as a reparameterization of
(\ref{eq1316}).

Consider a few
examples:

\subsection{Higgs potential}
The quartic potential for a complex scalar field  is given by
\begin{equation}
V_4(\Phi)=V_0 \pm m^2|\Phi|^2+\lambda|\Phi|^4
\label{eq1312}
\end{equation}
where the $-$ sign in front of the mass term gives the Higgs potential.
The potential may be written as
$V_4=V_0+V_{\pm}(|\Phi|^2/m^2)$ with
\begin{equation}
U_{\pm}(Y)\equiv \frac{1}{m^4} V_{\pm}(\eta Y)=\lambda\left( \eta Y\pm \frac{1}{2\lambda}\right)^2 -\frac{1}{4\lambda}\, .
\label{eq1212}
\end{equation}
Solving (\ref{eq1105}) we find
\begin{equation}
W_{\pm}(X)=\frac{1}{4\lambda}\left( \eta X\mp 1\right)^2\, .
  \label{eq1113}
\end{equation}

In the  example of ghost condensate
explored in \cite{ark1} the potential 
$W(X)=(  X -1)^2$ defined on the domain $X>1$ is TF equivalent to the canonical quartic potential
$V_+$ with $\eta=1$. The same potential on the
$X<1$ domain is TF equivalent to the phantom Higgs potential $V_-$ with
$\eta=-1$.

\subsection{Chaplygin gas}
In contrast to the standard assumption that dark matter and dark energy
are distinct, there stands the hypothesis that both are different
manifestations of a single entity. The first definite model of this
type \cite{kame5,bil4,fab16} is based on the Chaplygin
gas, an exotic fluid with an equation of state
\begin{equation}
p = - \frac{A}{\rho}
\label{eq000}
\end{equation}
Subsequently, the generalization to
\begin{equation}
p = - \frac{A}{\rho^\alpha}; \;\;\;\;\; 0 \leq \alpha \leq 1 
\label{eq500}
\end{equation}
was given
\cite{bent6} and the term `quartessence' coined \cite{mahl7}
to describe unified dark matter/energy models.

One of the most appealing aspects of the original Chaplygin gas model is that it is 
equivalent to the Dirac-Born-Infeld description of a D-brane in string theory.
\cite{bor,jac}. This may be seen as follows \cite{bil5}.
Consider a $p$-dimensional D-brane with coordinates $x^{\mu}$,
$\mu=0,1...p$,
 moving in the p+1-dimensional bulk with coordinates $X^{a}$,
$a=0,1...p+1$. 
In the string frame the action is given by \cite{john} 
\begin{equation}
S_{\rm DBI}= - \sqrt{A} \:
\int d^{p+1}x\, \sqrt{(-1)^p\det g^{(\rm ind)} }  \, 
\label{eq001}
\end{equation}
where $g_{\mu\nu}^{(\rm ind)}$ is the induced metric
or the ``pull back" of the bulk space-time metric 
$G_{ab}$ to the brane,
\begin{equation}
g^{(\rm ind)}_{\mu\nu}=G_{ab}
\frac{\partial X^a}{\partial x^\mu}
\frac{\partial X^b}{\partial x^\nu}\, .
\label{eq002}
\end{equation}

Let us 
 choose the coordinates such that $X^\mu=x^\mu$
and let the $p+1$-th coordinate $X^{p+1}\equiv\theta$  be normal to the brane.
From now on we set $p=3$ and consider a 3-brane universe in a 4+1 dimensional bulk.
Then
\begin{equation}
G_{\mu\nu}=g_{\mu\nu};
\;\;\;\;
\mu=0, ..., 3;\;\;\;\;
G_{\mu 4}=0; \;\;\;\;
G_{44}=-1\, .
\label{eq005}
\end{equation}
After a few straightforward algebraic manipulations,
the DBI action may be written as 
\begin{equation}
S_{\rm DBI} 
=\int d^4 x\sqrt{-\det g}\:
{\cal L}_{\rm DBI}\,; \;\;\;\;\;
{\cal L}_{\rm DBI}=-\sqrt{A}\sqrt{1-X}
\label{eq007}
\end{equation}
where $X$ is given by (\ref{eq1104}).
The energy-momentum tensor constructed from this Lagrangian
takes the perfect fluid form (\ref{eq510}) with
\begin{equation}
\rho=\frac{\sqrt{A}}{\sqrt{1-X}}
 \;\;\;\;\;\;\;\;
p=-\sqrt{A}\sqrt{1-X} 
\label{eq513}
\end{equation}
From this the Chaplygin gas equation. of state (\ref{eq000}) follows.

Using $W(X)={\cal L}_{\rm DBI}/m^4$, where $m^4=\sqrt{A}/2$,
we find the TF equivalent
\begin{equation}
U(Y)
 = Y+\frac{1}{Y} \, .
  \label{eq1107}
\end{equation}

\subsection{Symmetry breaking and condensation}
\label{symmetry}
Consider the
Lagrangian (\ref{eq1101})
for  a complex scalar field
with
 the potential $V(y)\geq 0$
  that possesses a minimum at some point
$y_0\equiv \phi_0^2/(2 m^2)$, i.e., 
assume that
there exists a point where $dV/dy=0$ and $d^2V/dy^2>0$.
In this case a spontaneous breakdown of U(1) symmetry of the
 Lagrangian (\ref{eq1101}) will take place.
 One of the components of the field $|\Phi|$ will have a nonzero vacuum
 expectation value such that the classical part of the field
 satisfies $|\Phi|^2=\phi_0^2/2$. There will be two
 quantum modes fluctuating around the minimum,
  one of them massive and the other
 one massless (Goldston boson).
 A typical   example is the Higgs potential
$V_4$ in (\ref{eq1312}) with
 the ``wrong" sign of the mass term.
The minima are obviously placed at positions satisfying $\phi_0^2 =m_0^2/\lambda$.
Another example is
the Chaplygin gas potential (\ref{eq1107})
\begin{equation}
V_{\rm Ch}=m^4\left(\frac{|\Phi|^2}{m^2} + \frac{m^2}{|\Phi|^2}\right) .
\label{eq2109}
\end{equation}
with the minima at $\Phi_0^2=m^2$.
In the neighborhoods of the minima,
this potential closely resembles (\ref{eq1312})
which may be seen by expanding (\ref{eq2109}) around
a minimum and identifying
$V_0\equiv 3 m^4$,
$m_0^2\equiv 2 m^2$, and $\lambda \equiv 1$.

In the neighborhood of a minimum, i.e., of a point $y_0$ 
where $dV/dy=0$ and $d^2V/dy^2>0$,
the solution will adiabatically role towards the minimum. Hence, we may assume 
that the solution
is almost static.
One finds three phases of the condensate corresponding to three types of solutions
according to the sign of $dV/dy$:

{\bf i)} $y>y_0$.
In this region $dV/dy>0$.
A configuration that solves
Eqs. (\ref{eq1204}) and (\ref{eq1206})
may be represented in terms of a self gravitating perfect fluid
with the 4-velocity
\begin{equation}
u^{\mu} = g^{\mu \nu}
 \theta_{,\nu}\left(\frac{1}{m^4}\frac{dV}{dy}\right)^{-1/2} \, ,
\label{eq2005}
\end{equation}
 and
with the equation of state given in a parametric form
(\ref{eq1216}).
In this case the dominant energy condition
$\rho \geq 0; \;\; 
\rho \pm p \geq 0$
 holds and
the corresponding set of solutions describe
a {\em canonical phase}.

The reduced Klein-Gordon equation
(\ref{eq1204}), (\ref{eq1206})
 may be further simplified
in the comoving reference frame, i.e., in the frame
 where the 4-velocity takes the form
$u^{\mu} =\delta^{\mu}_0/\sqrt{g_{00}}$.
 As a consequence of this and (\ref{eq2005})
the field $\theta$ is now a function of
$t$ only.
 Furthermore, Eq. (\ref{eq1206}) gives
\begin{equation}
\theta =\frac{\mu}{m} t + {\rm const}
\label{eq2202}
\end{equation}
where the constant $\mu$ is the
chemical potential 
associated to the conservation of U(1) charge \cite{lai,bil1}.
 Equation (\ref{eq1204}) degenerates now into an algebraic equation
\begin{equation}
\mu^2 g^{00}-\frac{1}{m^2} \frac{dV}{dy}=0
\label{eq2203}
\end{equation}
which relates the metric to the field $\phi$.
The quantity $\mu$ is introduced in the Euclidean path integral
by replacing
 the derivative
with respect to the Euclidean time $\tau=it$
with
\begin{equation}
\frac{\partial}{\partial \tau} \rightarrow
\frac{\partial}{\partial \tau}\pm \mu ,
\label{eq3203}
\end{equation}
where the $+$ or $-$ sign is taken when the derivative
acts on $\Phi^*$ or $\Phi$,
respectively.

{\bf ii)} $y<y_0$.
In this region $dV/dy<0$.
The procedure similar to i)
may be repeated for $\eta=-1$.
The velocity of the fluid is now defined as
\begin{equation}
u^{\mu} = g^{\mu \nu}
 \theta_{,\nu}\left(-\frac{1}{m^4}\frac{dV}{dy}\right)^{-1/2} \, ,
\label{eq2110}
\end{equation}
and we have a perfect fluid with
the EOS given by (\ref{eq1216}) as before
but $\rho+p<0$ so that the null and the dominant energy conditions
are now violated.
In the neighborhood of the minimum the field $\theta$ is again given by
(\ref{eq2202}) and the chemical potential
is defined by 
\begin{equation}
m^2\mu^2 g^{00}+\frac{ dV}{dy} =0
\label{eq2111}
\end{equation} 
The solutions to (\ref{eq1204}), (\ref{eq1206}) represent the
Bose-Einstein condensate of the  {\em phantom} field.
Hence, this type of solutions may be called
the {\em phantom phase}.

{\bf iii)} $y=y_0$.
This point is special. At this point
$dV/dy=0$ and equations
(\ref{eq1204}) and (\ref{eq1206}) yield a cosmological constant type
EOS
\begin{equation}
p=-\rho.
\label{eq2113}
\end{equation}
The equilibrium solutions  are configurations of constant density
$\rho=V(y_0)$.
This set of solutions describes the {\em de Sitter phase}.
Since the chemical potential $\mu$ is now equal to 0,
the condensation is due to the spontaneous symmetry braking
rather than due to an excess of positive or negative charge as in the case
of Bose-Einstein condensation at $\mu\neq 0$.


\section{Thermodynamics of purely kinetic k-essence}
We will base our thermodynamic analysis on a purely kinetic k-essence
 action 
\begin{equation}
S = \int \, d^{4}x \, \sqrt{- g}  \left[ - \frac{R}{16\pi G} + {\cal L} (X) \right],
\label{eq4001}
\end{equation}
with ${\cal L}$ in the form of (\ref{eq1203}). The Lagrangian ${\cal L}$ 
depends only on variable $X$ defined  in (\ref{eq1104})
with the field $\theta$ of dimension $m^{-1}$.
Such theories have been exploited as models for inflation
and dark 
matter/energy, e.g., purely kinetic k-essence \cite{arm1,arm2,sch1}
or ghost condensate \cite{ark1,kro1,kra1}.
A perfect fluid description applies for $X>0$.
Furthermore, equation (\ref{eq1106}) implies 
$\eta W_X>0$, which means that the domains where
$W_X>0$ correspond to a canonical scalar field Lagrangian 
(\ref{eq1101})
and those where $W_X<0$ to a phantom.
In particular, if in the neighborhood of $X=0$,
$W\sim \eta X$, then for $\eta=1$ the kinetic term
is  canonical
and  for $\eta=-1$ is of phantom type.
The field described by the Lagrangian (\ref{eq1203}) 
that behaves as a phantom near $X=0$
 is sometimes  called a ``ghost''.

The hydrodynamic quantities associated with eq. (\ref{eq4001}) are
\begin{equation}
p ={\cal L}\,; \;\;\;\;\; \rho = 2 X {\cal L}_{X}-{\cal L}. ,
\label{eq4003}
\end{equation}
and the fluid 4-velocity is given by (\ref{eq1215}.
Obviously, the  EOS defined
  by (\ref{eq4003})
in parametric form 
 is  barotropic. 
Next, we  start from  the standard thermodynamical relation
\begin{equation}
 d(\rho V)=
 T d S -pdV ,
\label{eq1601}
\end{equation}
where $V$ is the volume and $S$ is the entropy.
If there exist a conserved charge $Q$ with the corresponding
charge density $n$ the volume $V=1/n$, up to a constant factor. 
 Equation (\ref{eq1601}) may then be written in the form
\begin{equation}
 d\rho = T ds +\mu dn
\label{eq1603}
\end{equation}
where $s=S/V$ is the entropy density and we have introduced the chemical
potential
\begin{equation}
\mu= \frac{\rho+p-Ts}{n}
\label{eq1604}
\end{equation}
associated to the conserved charge $Q$.
In this way we define a grand-canonical ensemble in which
the thermodynamical quantities $p$, $\rho$ and $s$ 
are functions of two variables $\mu$ and $T$.
Equation (\ref{eq1604}) is nothing but the standard grand-canonical expression
for the entropy density
\begin{equation}
sT=p+\rho-\mu n.
\label{eq1605}
\end{equation}
Taking the derivative of this equation and combining it with 
(\ref{eq1603}) we find that the entropy and charge densities may be
expressed as partial derivatives of $p$
\begin{equation}
 s=\left.\frac{\partial p}{\partial T}\right|_\mu
\hspace{1cm}
n=\left.\frac{\partial p}{\partial \mu}\right|_T
\label{eq1607}
\end{equation}
Using this and (\ref{eq1605}) we find another
useful relation 
\begin{equation}
 p+\rho=T\frac{\partial p}{\partial T}+
\mu \frac{\partial p}{\partial \mu}
\label{eq1608}
\end{equation}
which may help us to narrow the arbitrariness in
functional dependence on $T$ and $\mu$.
It is clear from (\ref{eq1605}) that if $\mu=0$ the positivity of
entropy requires $p+\rho\geq 0$.
Hence, one could conclude that a phantom field must necessarily
yield a fluid with a negative entropy.
However, this conclusion is incorrect since
generally $\mu\neq 0$ and the entropy density given by
(\ref{eq1605}) need not be negative.
In fact, as we will shortly demonstrate,
for an arbitrary temperature $T$
it is always possible to find a range of $\mu$ such
that $s\geq 0$.

From (\ref{eq4003}) and
(\ref{eq1608}) it follows that
the variable $X$ as a function of $T$ and $\mu$
satisfies a partial differential equation
\begin{equation}
 T\frac{\partial X}{\partial T}+
\mu \frac{\partial X}{\partial \mu}=2X.
\label{eq4005}
\end{equation}
The most general solution to this equation is 
a homogeneous function of 2nd degree
which may be written as
\begin{equation}
X= \frac{\mu^2}{m^2} f(T/\mu)
\label{eq4006}
\end{equation}
or equivalently
\begin{equation}
X= \frac{T^2}{m^2} g(\mu/T)\, .
\label{eq4007}
\end{equation}
Here $f$ is an arbitrary positive dimensionless function of $x\equiv T/\mu$
and the function $g$ is related to $f$  by
$f(x)= x^2 g(1/x)$.
%
The entropy density  may be calculated from (\ref{eq1607}).
With (\ref{eq4003}) and (\ref{eq4007}) we find
\begin{equation}
s= X{\cal L}_X \frac{1}{\mu} \frac{f'}{f}
\label{eq4009}
\end{equation}
Now we require $S> 0$ at $T\neq 0$ and $S=0$ at $T=0$.
From the latter requirement it follows $f'(0)=0$.
The  requirement $S\geq 0$ and the condition
$\eta {\cal L}_X>0$
 imply
$\eta\mu f' > 0$ for $T\neq 0$.
A simple nontrivial function that satisfies the above conditions
is, e.g.,
\begin{equation}
f=C_1+\eta C_2 x^2
\label{eq4010}
\end{equation}
where $C_1$ and $C_2$ are arbitrary positive constants.
However, in the phantom case the positivity of $f$ puts 
an additional constraint on $\mu$ and $T$:
\begin{equation}
\frac{T}{\mu}\leq \frac{C_1}{C_2}
\label{eq4011}
\end{equation}

\section{Chemical potential}
\label{chemical}
The equation of motion (\ref{eq1304})
for the field $\theta$ 
is in fact a  conservation equation for
the current
\begin{equation}
j^{\mu}=2m^3 W_X g^{\mu\nu}\theta_{,\nu}\, .
\label{eq3001}
\end{equation}
The current conservation is related to the symmetry under the constant shift 
 $\theta\rightarrow \theta+c$ of the scalar field $\theta$.
The conserved charge is defined as
\begin{equation}
Q=\int_{\Sigma} j^\mu d\Sigma_{\mu}
=\int_{\Sigma} n\, u^{\mu}d\Sigma_{\mu}
\label{eq3002}
\end{equation}
where the integration goes over an arbitrary spacelike hypersurface $\Sigma$
 that contains
the charge.
Using the definition (\ref{eq1215}) for the velocity, we obtain the charge density as
\begin{equation}
n= 2 m^3 \sqrt{X} W_X 
\label{eq3103}
\end{equation} 
In the Hamiltonian formulation \cite{wal} we choose
the hypersurface $\Sigma$ at constant time so that the total charge
(\ref{eq3002}) becomes a volume integral
\begin{equation}
Q= 2 m^3 \int_V  W_X g^{0\nu}\theta_{,\nu}dV
\label{eq3003}
\end{equation}
The current (\ref{eq3001}) corresponds
to the Klein-Gordon current
\begin{equation}
j_{\rm KG}^{\mu}=ig^{\mu\nu}(\Phi^*\Phi_{,\nu}-\Phi\Phi^*_{,\nu}),
\label{eq3004}
\end{equation}
which is the conserved current in the TF equivalent theory.

Next, we introduce the chemical potential $\mu$ associated to the conserved
 charge (\ref{eq3002}).
To find the effective Lagrangian that contains the chemical potential
we start from the grandcanonical partition function
\begin{equation}
Z=\mbox{Tr}\; e^{-\beta (\hat{H}-\mu \hat{Q})}=\int [d\pi]
\int_{\rm periodic} [d\theta]\exp \int_0^\beta d\tau \int dV \left( i\pi 
\frac{\partial\theta}{\partial\tau}
-{\cal H} +\frac{\mu}{m}\pi\right)\, ,
\label{eq3005}
\end{equation}
where $\beta=1/T$, $\pi$ is the conjugate momentum field and the 
 Hamiltonian  density ${\cal H}$ is related to ${\cal L}$ by the 
usual Legendre transformation. 
A formal functional integration of (\ref{eq3005}) over $\pi$ yields the partition function 
expressed in terms of the effective Euclidean Lagrangian
\begin{equation}
Z=
\int_{\rm periodic} [d\theta]\exp -\int_0^\beta d\tau \int dV 
{\cal L}_{\rm E}(\theta, \mu)\, .
\label{eq3009}
\end{equation}
Unfortunately, the analytic functional integration is generally not possible. 
Nevertheless, in the in the saddle point
approximation we find
\begin{equation}
-{\cal L}_{\rm E}= {\cal L}( i\frac{\partial\theta}{\partial\tau}
+\frac{\mu}{m},\theta_{,i}) 
\label{eq3012}
\end{equation}
Hence, the effective Lagrangian is obtained from
the Lagrangian (\ref{eq1203})
by replacing the derivatives of the field $\theta$ by
\begin{equation}
\theta_{,\nu}\rightarrow
\theta_{,\nu}+\frac{\mu}{m}\delta^0_\nu\, .
\label{eq3013}
\end{equation}
Note the difference and similarity with the Euclidean field theory
prescription
(\ref{eq3203})
for a canonical complex scalar field.

We now check the consistency of the prescription (\ref{eq3013}) with the solution
(\ref{eq4006}). To do that, it is useful to work in the comoving reference frame
i.e., in the frame
 where the 4-velocity takes the form
$u^{\mu} =\delta^{\mu}_0/\sqrt{g_{00}}$. Comparing this with the definition of 
the 4-velocity (\ref{eq1215}) we conclude that $\theta$ is a function of $t$ only.
Then
\begin{equation}
X=g^{00}
(\theta_{,0}+\mu/m)^2.
\label{eq3014}
\end{equation}
This compared with the general expression
(\ref{eq4006}) implies $\theta_{,0}=0$, $T=0$, and $S=0$.
Hence, we conclude that the consistency of (\ref{eq3013}) with 
(\ref{eq4006}) implies zero temperature and zero entropy for a general
purely kinetic k-essence type of theory.
 
Two remarks are in order. First, one should bear in mind that this result is obtained 
using the
effective Euclidean Lagrangian (\ref{eq3012}) derived in a saddle point approximation.
Second, the solution (\ref{eq4006}) is classical whereas 
the partition function
(\ref{eq3009}) generally represents quantum and thermal fluctuations
of the field $\theta$.
\section{Summary and conclusion}
\label{conclusion}
Using the Thomas Fermi correspondence we have shown that a general
DE model based on a complex scalar field theory can be 
equivalently represented by a purely kinetic k-essence modell. 
Our thermodynamic analysis of purely kinetic k-essence shows
that the entropy is positive or zero and need not be negative
for phantom theories contrary to the claims
often stated in the recent literature
(see, e.g., \cite{gong} and references therein)
that a violation of
NEC  implies
negative entropy.
Furthermore, using the grandcanonical partition function
derived in a saddle point approximation of the functional integral
over conjugate momenta and comparing the effective
Euclidean Lagrangian with the quintessence equation of state
we obtain quite generally the chemical potential $\mu\neq 0$
but the temperature $T=0$ and the entropy  $S=0$.
In principle,  nontrivial thermal contributions can be obtained 
from
quadratic fluctuations of the field around the classical solution.
However this would go beyond the scope of this paper and will be done elswhere.

\subsection*{Acknowledgments}
This work was supported by the Ministry of Science,
Education and Sport
of the Republic of Croatia under contract No. 098-0982930-2864 and
partially supported through the Agreement between the Astrophysical
Sector, S.I.S.S.A., and the Particle Physics and Cosmology Group, RBI.




\begin{thebibliography}{99}

%
\bibitem{bre}
  I.~Brevik, S.~Nojiri, S.~D.~Odintsov and L.~Vanzo,
  Phys.\ Rev.\  D {\bf 70} (2004) 043520
  [arXiv:hep-th/0401073];
  S.~Nojiri and S.~D.~Odintsov,
  Phys.\ Rev.\  D {\bf 70} (2004) 103522
  [arXiv:hep-th/0408170].
\bibitem{lim}
  J.~A.~S.~Lima and J.~S.~Alcaniz,
  Phys.\ Lett.\  B {\bf 600} (2004) 191
  [arXiv:astro-ph/0402265].
\bibitem{gonz}
  P.~F.~Gonzalez-Diaz and C.~L.~Siguenza,
  Nucl.\ Phys.\  B {\bf 697} (2004) 363
  [arXiv:astro-ph/0407421].
\bibitem{izq}
  G.~Izquierdo and D.~Pavon,
  Phys.\ Lett.\  B {\bf 633} (2006) 420
  [arXiv:astro-ph/0505601].
\bibitem{moh}
  H.~Mohseni Sadjadi,
  Phys.\ Rev.\  D {\bf 73} (2006) 063525
  [arXiv:gr-qc/0512140].
\bibitem{set}
  M.~R.~Setare and S.~Shafei,
  JCAP {\bf 0609} (2006) 011
  [arXiv:gr-qc/0606103];
  M.~R.~Setare,
  Phys.\ Lett.\  B {\bf 641} (2006) 130
  [arXiv:hep-th/0611165].
%
\bibitem{wan}
  B.~Wang, Y.~Gong and E.~Abdalla,
  Phys.\ Rev.\  D {\bf 74} (2006) 083520
  [arXiv:gr-qc/0511051].
\bibitem{gong}
  Y.~Gong, B.~Wang and A.~Wang,
  Phys.\ Rev.\  D {\bf 75} (2007) 123516
  [arXiv:gr-qc/0611155].
\bibitem{san}
  F.~C.~Santos, M.~L.~Bedran and V.~Soares,
  Phys.\ Lett.\  B {\bf 636} (2006) 86;
  Phys.\ Lett.\  B {\bf 646} (2007) 215.
\bibitem{peeb8}
C.\ Wetterich, Nucl.\ Phys.\ B {\bf 302}, 668 (1988);
P.J.E.\ Peebles and B.\ Ratra,
Astrophys.\ J.\ {\bf 325}, L17 (1988).
\bibitem{arm2}
  C.~Armendariz-Picon, V.~F.~Mukhanov and P.~J.~Steinhardt,
  Phys.\ Rev.\ Lett.\  {\bf 85} (2000) 4438
 \bibitem{cop}
  E.~J.~Copeland, M.~Sami and S.~Tsujikawa,
  Int.\ J.\ Mod.\ Phys.\  D {\bf 15} (2006) 1753
  [arXiv:hep-th/0603057].
%
\bibitem{par18}
The terminology is borrowed from Bose-Einstein condensates:
A.S.\ Parkins and D.F.\ Walls,  Phys.\ Rep.\ {\bf 303}, 1 (1998).
%
\bibitem{ark1}
 N.\ Arkani-Hamed, H.C.\ Cheng, M.A.\ Luty, and S.\ Mukohyama,
 JHEP {\bf 05} (2004) 074.
%
\bibitem{bil4} N.\ Bili\'{c}, G.B.\ Tupper, and R.D.\ Viollier,
 Phys.\ Lett.\ B{\bf 535} (2002) 17.
%
\bibitem{kame5} A.\ Kamenshchik, U.\ Moschella, and V.\ Pasquier,  
Phys.\ Lett.\ B{\bf 511} (2001) 265.
%
\bibitem{fab16} 
  J.~C.~Fabris, S.~V.~B.~Goncalves and P.~E.~De Souza,
  Gen.\ Rel.\ Grav.\  {\bf 34} (2002) 53;
  Gen.\ Rel.\ Grav.\  {\bf 34} (2002) 2111.
\bibitem{bent6} M.C.\ Bento, O.\ Bertolami, and A.A.\ Sen,
 Phys.\ Rev.\ D{\bf 66} (2002) 043507.
%
\bibitem{mahl7} M.\ Makler, S.Q.\ de Oliveira, 
and I.\ Waga,  Phys.\ Lett.\ B{\bf 555} (2003) 1.
%
\bibitem{bor}
M.\ Bordemann and J.\ Hoppe, Phys. Lett. B{\bf 325} (1994) 359;
N.\ Ogawa, Phys.\ Rev. D{\bf 62} (2000) 085023.
\bibitem{jac}
R. Jackiw, {\it Lectures on Fluid Mechanics} (Springer Verlag, Berlin, 2002).
%
\bibitem{bil5}
N.\ Bili\'{c}, G.B.\ Tupper, and R.D.\ Viollier,
J.\ Phys.\ A {\bf 40} (2007) 6877; gr-qc/0610104.
\bibitem{john}
C.V. Johnson, D-Branes,  Cambridge University Press, Cambridge, 2003.
%
\bibitem{lai}
M. Laine and M. Shaposhnikov,
Nucl. Phys. B 532 (1998) 376.
%
\bibitem{bil1}
N. Bili\'c and H. Nikoli\'c,
Nucl Phys. B 590 (2000) 575
%
%
\bibitem{arm1}
  C.~Armendariz-Picon, T.~Damour and V.~F.~Mukhanov,
  Phys.\ Lett.\ B {\bf 458} (1999) 209
  [arXiv:hep-th/9904075].
\bibitem{sch1}
 R.J.\ Scherrer, Phys.\ Rev.\ Lett.\ {\bf 93} (2004) 011301.
%
\bibitem{kro1}
  D.~Krotov, C.~Rebbi, V.~A.~Rubakov and V.~Zakharov,
  Phys.\ Rev.\ D {\bf 71} (2005) 045014,
  arXiv:hep-ph/0407081.
\bibitem{kra1} 
A.~Krause and S.~P.~Ng,
  Int.\ J.\ Mod.\ Phys.\ A {\bf 21} (2006) 1091,
 arXiv:hep-th/0409241.
%
%
\bibitem{wal}
R.M. Wald,
{\em General relativity},
University of Chicago, Chicago 1984.
 \end{thebibliography}
\end{document}